# On the role of long range internal stresses on grain nucleation during discontinuous recrystallization

Paul DUVAL, François LOUCHET*, Jérôme WEISS and Maurine MONTAGNAT

Laboratoire de Glaciologie et de Géophysique de l'Environnement,UJF/CNRS,
BP96, F-38402 St Martin d'Hères cedex, France
*corresponding author: louchet@lgge.obs.ujf-grenoble.fr

**Abstract:** The essential role of long range elastic interactions in recrystallization is demonstrated using a simple analytical model: pileup rearrangements following absorption of leading dislocations by a dislocation-free embryo provides an additional driving force that results in a drastic decrease of both the nucleation critical radius and the saddle point energy. A very sharp transition is evidenced, at which the saddle point totally disappears and nucleation becomes spontaneous. This transition occurs for a well defined critical stress corresponding to both a critical density of geometrically necessary dislocations and a critical strain, without invoking any critical nucleus size that may be reached with the help of some dislocation microstructure instability. The present model is illustrated here by the case of polycrystalline ice, but may apply to other crystalline material with significant plastic anisotropy, as Zircaloy for instance.

## 1. Introduction

Dynamic recrystallization at relatively high temperatures may take place during deformation. It can occur through several mechanisms and can be continuous or discontinuous (1). During continuous recrystallization, sub-boundaries form as deformation proceeds, and high-angle boundaries may result from progressive misorientation of sub-boundaries. In this recrystallization regime, grain boundaries migrate in the same low-velocity regime as in normal grain growth (2). By contrast, discontinuous recrystallization results from rapid migration of grain boundaries that separate dislocation-free nuclei from deformed grains. The migration rate is much larger than that associated with continuous recrystallization (2,3).

The nucleation critical radius is usually derived considering the balance between the stored elastic energy within the nucleus (associated with the total dislocation energy) and the interfacial energy of the boundary (4, 5, 6). Taking the example of ice, the critical radius would be of a few mm and the saddle point energy of about $10^{-4}$ J = 7 $10^{14}$ eV, making a thermally activated nucleation totally unrealistic. The situation is similar for other crystalline materials.

The aim of the present work is to analyse nucleation processes during discontinuous dynamic recrystallization by considering the interaction of a nucleating grain with the long-range internal stress field that develops during primary creep, mainly in crystals subjected to a large plastic anisotropy, as ice or Zircaloy for instance.

We show here that taking into account the relaxation of long-range internal stress fields during nucleation drastically changes the situation. We estimate the role of long range stresses involved in the recrystallisation process considering pileup relaxation during the embryo nucleation process, and discuss the consequences on recrystallization kinetics. The present analysis is developed taking ice as a model material, characterised by a significant plastic anisotropy: a strong Bauschinger effect is indeed observed in this material (7), showing that the geometrically necessary dislocation (GND) density $\rho_{gnd}$ should be at least one order of magnitude larger than the statistical dislocation density $\rho_s$.

## 2. Change in stored elastic energy associated with nucleation:

As dislocation stress fields are long-ranged, the nucleation of a dislocation-free embryo modifies the dislocation structure and the associated stress field up to very large distances. This effect is particularly enhanced for GNDs that accommodate plastic heterogeneities induced by strain incompatibility between grains. In order to estimate the corresponding energy reduction, we mimick heterogeneous GND structures by a series of pileups. We consider the nucleation of a dislocation-free embryo on a grain boundary, at the head of a pileup of length *L* (of the order of the grain size) containing *n* dislocations labelled *1* to *n*, with separations $x_1, x_2, ..., x_p$ measured from the pileup head. These quantities are related by (8):

$$x_p = \frac{np}{n-p} x_1, \quad x_1 = \frac{\mu b}{4n\sigma}, \quad n = \frac{2\sigma L}{\mu b} \quad (1)$$

where $\mu$ is the shear modulus, *b* the Burgers vector length, and $\sigma$ the resolved applied stress in the considered slip system.

Let us now estimate the energy provided by a reduction of the number of dislocations at the pileup head due to the embryo nucleation. The number of dislocations of a pileup contained in a sphere of radius *r* centered on the pileup tip is obtained setting $x_p = r$, which gives after some simple algebra:

$$p = \frac{n}{1+\dfrac{\mu b}{4\sigma r}} = n\frac{1}{1+\xi} \quad (2)$$

with:

$$\xi = \mu b / 4\sigma r \quad (3)$$

If those *p* dislocations are swept out during nucleation of an embryo of radius *r*, the corresponding gain of energy can be expressed as the energy difference between a pileup with *n* dislocations and a pileup with *(n-p)* dislocations:

$$\Delta w = -\frac{\pi(1-\nu)}{8}\mu b^2 \left[n^2 - (n-p)^2\right] = -\frac{\pi(1-\nu)}{8}\mu b^2 p(2n-p) \quad (4)$$

with :

$$2n - p = n\left(2 - \frac{1}{1+\xi}\right) = n\frac{1+2\xi}{1+\xi} \quad (5)$$

which gives, using eqs. (1) to (5) :

$$\Delta w = -\frac{\pi(1-\nu)}{2}\frac{\sigma^2 L^2}{\mu}\frac{1+2\xi}{(1+\xi)^2} \quad (6)$$

We also have to consider two points:

i) eq. (6) is valid for unit dislocation length. We shall consider that the pileup is rearranged over a transversal length (perpendicular to pileup direction) *2αr*, where *α* is a scaling factor of the order of (but probably larger than) unity. The relaxed energy given by eq. (6) has to be multiplied by this scaling factor.

ii) considering large enough embryo radii and/or GND densities, several pileups may interact with a single embryo. If $\rho_{gnd}$ is the average GND density, the separation between parallel pileup planes is of the order of $\rho_{gnd}^{-1/2}$. We therefore consider that the embryo of diameter *2r* intercepts a number of pileups of the order of $2r\rho_{gnd}^{1/2}$.

We consider the situation where a given stress value results in a saturation strain for which the applied stress is balanced by pileup back stresses. The GND density is therefore directly related to the applied stress, and the number $2r\rho_{gnd}^{1/2}$ of pileups intercepted by the embryo can be parametrized as a function of stress: considering that the total number $\rho_{gnd} L^2$ of GNDs in the grain is equal to the number $N_{pu}$ of pileups times the number *n* of dislocations per pileup, we can write:

$$\rho_{gnd} L^2 = n N_{pu} = n \frac{L}{1/\sqrt{\rho_{gnd}}} = nL\sqrt{\rho_{gnd}} \tag{7}$$

Replacing *n* by its expression given in eq. (1) gives:

$$\sqrt{\rho_{gnd}} = \frac{2\sigma}{\mu b} \tag{8}$$

which is equivalent to Taylor's relation (9).

Taking into account points i) and ii), and using eq. (8), the total decrease in energy associated to GND relaxation is given by:

$$\Delta W_{gnd} = -\frac{4\pi\alpha(1-\nu)}{\mu^2 b}\sigma^3 L^2 \frac{1+\mu b/2\sigma r}{(1+\mu b/4\sigma r)^2} r^2 \tag{9}$$

In addition, the classical terms for bulk and interface energies can be written:

$$\Delta W_{bulk} = -\frac{4}{3}\pi\rho_s \mu b^2 r^3 \tag{10}$$

where $\rho_s$ is the statistical dislocation density only, since GNDs are taken into account in eq. (9), and:

$$\Delta W_{gb} = 4\pi\gamma_{gb} r^2 \tag{11}$$

where $\gamma_{gb}$ is the grain boundary energy.

The total energy balance taking into account the classical terms and the additional term *ΔW$_{gnd}$* is therefore:

$$\Delta W_{tot} = -\frac{4}{3}\pi\rho_s \mu b^2 r^3 + \left[-\frac{4\pi\alpha(1-\nu)}{\mu^2 b}\sigma^3 L^2 \frac{1+\mu b/2\sigma r}{(1+\mu b/4\sigma r)^2} + 4\pi\gamma_{gb}\right]r^2 \tag{12}$$

Eq. (12) holds as long as the average number of pileups intercepted by the embryo is larger than *1*. Otherwise, nucleation is expected to occur at pileup tips only, and not at grain boundary sites that are not intercepted by a pileup. In this case, the number of pileups intercepted by the embryo, taken of the order of $2r\rho_{gnd}^{1/2}$ in eq. (12), has to be replaced by *1*. Using eq. (8), the condition $2r\rho_{gnd}^{1/2} \ll 1$ also writes *μb/4σr >>1*, and *1* can be neglected in front of *μb/4σr* in eq. (12), which gives:

$$\Delta W_{tot} \approx -\frac{4}{3}\pi \rho_s \mu b^2 r^3 + \left[ -\frac{8\pi\alpha(1-\nu)}{\mu^2 b}\sigma^3 L^2 + 4\pi\gamma_{gb}\right] r^2 = -Ar^3 + [-B+C]r^2 \qquad (13)$$

where *A* and *C* are the coefficients of the classical volume and interface terms, and where *B* is associated with GNDs, i.e. long range internal stresses.

### 3. Results and discussion:

The following results show the variations of *ΔW*$_{tot}$ for ice, as a function of embryo radius *r*, for different stress values, and obtained from eq. (12) in which *L* is taken of the order of the grain size. Parameters values are as follows: µ=3 GPa, ν=0.33, b=4.5 10$^{-10}$m, γ$_{GB}$=0.065 J m$^{-2}$, ρ$_s$=10$^{10}$ m$^{-2}$, α=1, L=5 mm (10).

The classical case is obtained setting to zero the first term in the bracket in eq. (12). Despite the fact that $\rho_s$ has been given a value (10$^{10}$ m$^{-2}$) larger than what is usually observed in ice, in order to take into account the total dislocation density as usually considered, a huge saddle point energy *W*\* is obtained (more than 10$^{-4}$ joule = 7 10$^{14}$ eV), to be compared to *kT*=1/40 eV at room temperature, associated with a very large critical radius (2cm). As usually agreed, such values obviously preclude any possibility of thermally activated nucleation of a critical embryo within a reasonable time. Decreasing *W*\* down to 1eV, i.e. down to a value allowing thermal activated nucleation, would require a dislocation density of about 10$^{17}$ m$^{-2}$, i.e. dislocation separations of the order of 1 nm, never observed in crystalline materials. As a consequence, ignoring long range stresses associated with GNDs cannot account for observations.

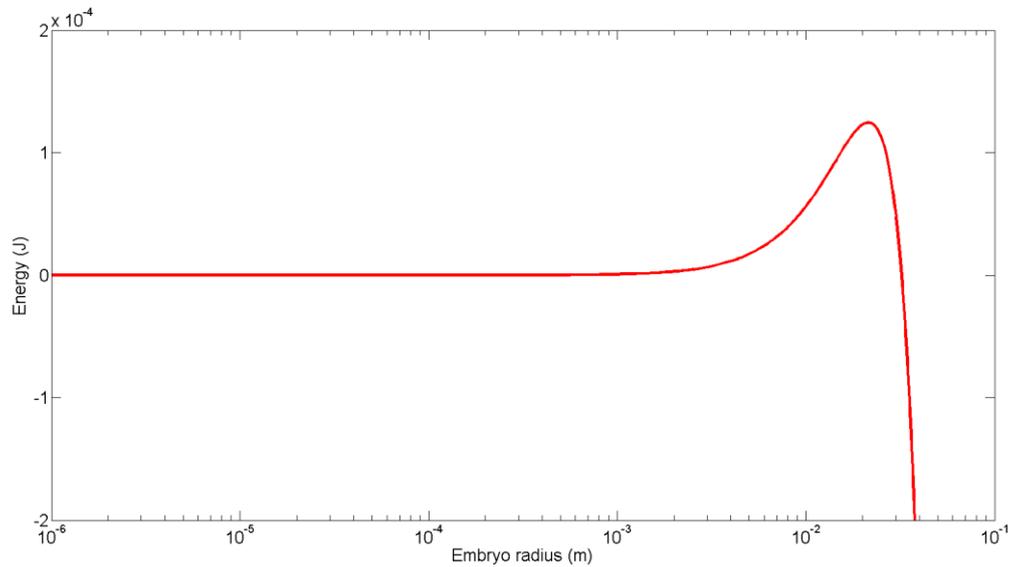

Fig. 1

Variations of the total embryo energy as a function of embryo radius in the classical case (no long range internal stresses).

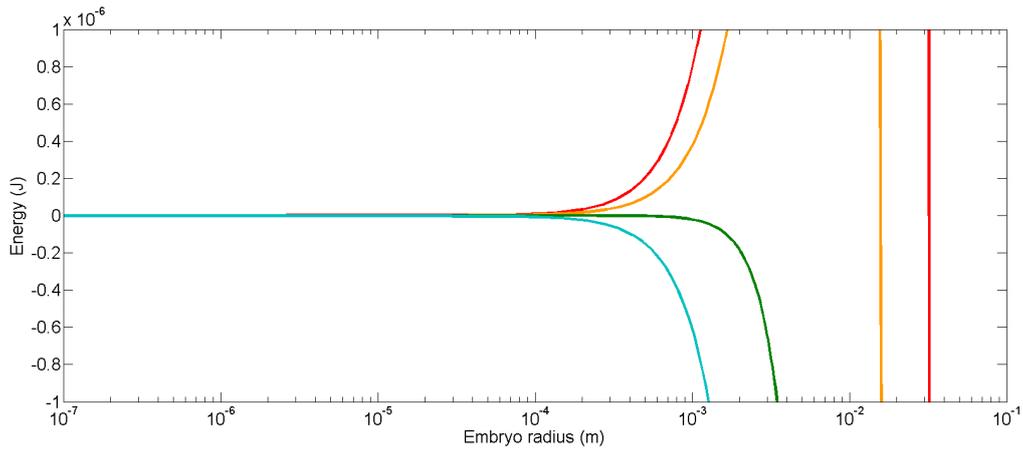

Fig. 2

Variations of the total embryo energy as a function of embryo radius, for a grain size of 5 mm, and different stress values, corresponding to different $\rho_{gnd}$ values: classical case (red), σ=0.020 MPa (orange), σ=0.025 MPa (green), σ=0.030 MPa (blue). The transition between the saddle point behaviour and the spontaneous nucleation occurs between 0.020 and 0.025 MPa.

Figs. 2 and 3 illustrate the behaviour of the system under increasing stress values. They are obtained from eq. (12) where $2r\rho_{gnd}^{1/2} = 4\sigma r/\mu b$ has been replaced by *1* every time it was smaller than *1*. As seen in Fig. 2, introducing some GND density through an increase of stress drastically decreases the saddle point energy *W\**. Even more interesting, the *W\*(r)* curve exhibits a very sharp transition from a saddle point configuration to a continuously decreasing curve that corresponds to a spontaneous nucleation. The transition takes place for stresses between 0.020 MPa (for which the saddle point energy is still huge) and 0.030 MPa (for which nucleation is spontaneous).

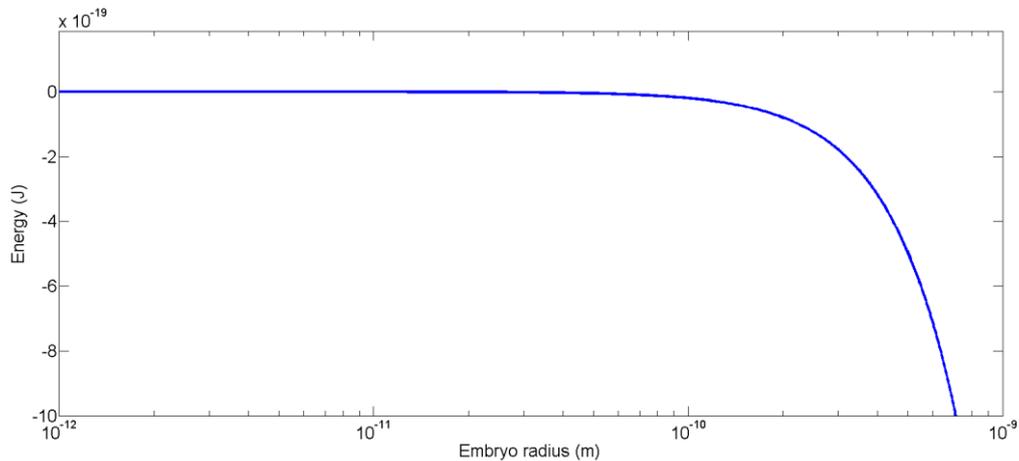

Fig. 3

Details of the total embryo energy vs embryo radius at small energies and radii, for an applied stress σ=0.030MPa: no saddle point is visible.

Fig. 3 zooms the *W(r)* curve down to very small energy values of the order of the electron volt, showing that even at such a scale, there is no visible saddle point. This finding can be confirmed analytically from eq. (13), valid for small *r* values. A saddle point exists only if *B<C*, otherwise *W* is a uniformly decreasing function of *r*. The critical stress $\sigma_c$ corresponding to the transition is therefore given by *B=C*, i.e.:

$$\sigma_c = \left[\frac{\mu^2 b \gamma_{gb}}{2\alpha(1-\nu)L^2}\right]^{1/3} \qquad (14)$$

Using the same numerical values as above, we find $\sigma_c$ =0.02 MPa, in very good agreement with numerical results of fig. 3 ($\sigma_c$ around 0.025 MPa).

The critical strain can be clearly defined as the strain for which the saddle point configuration turns into a spontaneous athermal behaviour, which gives:

$$\varepsilon_c \approx \rho_c bL = \left(\frac{2\sigma_c}{\mu b}\right)^2 bL = \frac{4\sigma_c^2 L}{\mu^2 b} \qquad (15)$$

where $\rho_c$ is the critical value of $\rho_{gnd}$ corresponding to the critical stress $\sigma_c$. Taking $\sigma_c$ = 0.025 MPa as suggested by fig. 2, we find a critical strain $\varepsilon_c$ of about 0.3%, in reasonable agreement with experiment (11).

Such a sharp transition between a huge saddle point and a continously decreasing embryo energy means that thermally activated nucleation should not be considered at all. The role of temperature restricts to its influence on grain boundary migration rate only.

### 4. Summary and conclusion:

The present model shows the essential role of long range internal stresses in grain nucleation during discontinuous recrystallization in ice, schematized by pile-up relaxation during embryo formation. The relaxation of dislocation pile-ups contributes to a significant decrease of both the critical radius and the saddle point energy, leading to a very sharp transition for a critical stress value: at the transition, the saddle point totally vanishes, making nucleation spontaneous, without invoking any critical nucleus size that may be reached with the help of some dislocation microstructure instability (5).

The critical strain $\varepsilon_c$ can be clearly defined as the strain for which the saddle point disappears, and is directly related to the presence of large long-range internal stresses. It is worth noting that a similar role of long-range internal stresses was also shown to be significant during martensitic transformations (12).

As no thermal activation is necessary in the nucleation process, the influence of temperature is restricted to grain boundary migration kinetics, making rapid migration of grain boundaries an obvious prerequisite to discontinuous recrystallization (2, 3). Since embryo nucleation modifies the dislocation structure and the associated stress field up to very large distances, such an athermal event may possibly trigger nucleation cascades up to significant distances.

The present model is illustrated here by the case of polycrystalline ice. In spite of its crudeness, the critical stress and strain are in very good agreement with observations. This model may obviously apply to other crystalline materials with significant plastic anisotropy, as Zircaloy for instance.

Potential consequences of such spontaneous processes open a new and innovative investigation area.

**Acknowledgements:**

P. Duval and F. Louchet are grateful to the "Centre National de la Recherche Scientifique" and the "Institut National Polytechnique de Grenoble" for providing them emeritus positions, and making available to them the facilities of the LGGE department.

**References:**

(1) M. Guillopé, and J.P. Poirier, Dynamic recrystallization during creep of singlecrystalline halite : an experimental study. *J. of Geophys. Res*. B10, **84**, 5557 (1979).

(2) J.P. Poirier, *Creep of crystals; high temperature deformation processes in metals, ceramics and minerals*. Cambridge: Cambridge University Press (1985).

(3) P. Duval, and O. Castelnau, Dynamic recrystallization of ice in polar ice sheets. *Journal de Physique IV*, 5 (Colloque N°3), **C3**-197- 205 (1995).

(4) J.W. Christian, *The theory of transformations in metals and alloys*. Pergamon Press (1965).

(5) F.J. Humphreys, and M. Hatherly, *Recrystallization and related annealing phenomena*. Oxford: Pergamon (1996).

(6) Y. Bréchet, and G. Martin, Nucleation problems in metallurgy of the solid state: recent developments and open questions. *C.R. Physique*, **7**, 959-976 (2006).

(7) P. Duval, M.F. Ashby, and I. Anderman, Rate-controlling processes in the creep of polycrystalline ice, *J. of Phys. Chemistry*, 87, 4066-4074 (1983).

(8) J. Friedel, *Dislocations*, Pergamon, Oxford, p. 261-262 (1964).

(9) G. I. Taylor, J. Inst. Met. **62**, 307 (1938).

(10) E.M. Schulson, and P. Duval, *Creep and fracture of ice* Cambridge University Press (2009).

(11) T.H. Jacka, The time and strain required for development of minimum strain rates in ice, *Cold Reg. Sc. Techno.*, 8, 261-268 (1984).

(12) Q.P. Meng, Y.H. Rong, and T.Y. Hsu, Effect of internal stress on autocatalytic nucleation of martensitic transformation. *Metall. and Mater. Transactions*. **37A**, 1405-1411 (2006).